\documentclass[lettersize,journal]{IEEEtran}
\usepackage{amsmath,amsfonts,amssymb,gensymb}
\usepackage[noend]{algorithmic}
\usepackage{algorithm}
\usepackage{array}
\usepackage[caption=false,font=normalsize,labelfont=sf,textfont=sf]{subfig}
\usepackage{textcomp}
\usepackage{stfloats}
\usepackage{url}
\usepackage{verbatim}
\usepackage{graphicx}
\usepackage{cite}
\usepackage{booktabs} 
\usepackage{multirow}
\usepackage{xcolor}
\usepackage[super]{nth}
\usepackage{tikz}
\newcommand*\circled[1]{\tikz[baseline=(char.base)]{
            \node[shape=circle,draw,inner sep=0.2pt] (char) {#1};}}
\newcommand*\circledB[1]{\tikz[baseline=(char.base)]{
            \node[shape=circle,fill,inner sep=0.2pt] (char) {\textcolor{white}{#1}};}}
\usetikzlibrary{tikzmark}
\tikzset{circledColor/.style={circle,draw,inner sep=0.1em,line width=0.04em}}
\usepackage{soul}
\usepackage{enumitem}

\usepackage[switch]{lineno}

\begin{document}


\title{DxPTA: An Architecture Design Space Exploration with Optical Dataflow-guided Strategy for HW/SW Co-Design of Photonic Transformer Accelerators}

\author{Rachmad Vidya Wicaksana Putra,~\IEEEmembership{Member,~IEEE,} Solomon Micheal Serunjogi, Mahmoud Rasras,~\IEEEmembership{Senior Member,~IEEE,} and Muhammad Shafique,~\IEEEmembership{Senior Member,~IEEE} 
\thanks{Rachmad Vidya Wicaksana Putra is with eBRAIN Lab, Division of Engineering, New York University (NYU) Abu Dhabi, United Arab Emirates; 
(e-mail: rachmad.putra@nyu.edu). \\
\indent Solomon Micheal Serunjogi is with Photonic Research Lab (PRL), Division of Engineering, New York University (NYU) Abu Dhabi, United Arab Emirates; 
(e-mail: sms10215@nyu.edu). \\
\indent Mahmoud Rasras is the Director of Photonic Research Lab (PRL), Division of Engineering, New York University (NYU) Abu Dhabi, United Arab Emirates (UAE); 
(e-mail: mrasras@nyu.edu). \\
\indent Muhammad Shafique is the Director of eBRAIN Lab, Division of Engineering, New York University (NYU) Abu Dhabi, United Arab Emirates; 
(e-mail: muhammad.shafique@nyu.edu). }
}


\maketitle


\begin{abstract}
Transformer-based networks have emerged as prominent AI models with state-of-the-art performance, which potentially pave the way toward artificial general intelligence (AGI). 
However, their large sizes still hinder their efficient implementation, thus highlighting the need for alternate solutions to enable their energy-efficient acceleration. 
Recently, state-of-the-art works propose photonic transformer accelerators (PTAs) with significant speedup and energy efficiency improvements over the conventional electronic accelerators.
However, their PTA architectures are developed without considering the application constraints (e.g., area, power, energy, and latency).
Moreover, their manual design approach also requires huge design time to determine a suitable architecture for the targeted application, hence making this approach not scalable.
To address these limitations, we propose \textit{\textbf{DxPTA}}, a novel design space exploration methodology for enabling efficient hardware/software co-design of the appropriate PTA architecture that meets all constraints. 
It is achieved by (1) identifying the PTA architecture parameters based on the coherent optical dataflow; (2) analyzing the impact/significance of the parameters; and (3) leveraging this analysis for devising a constraint-aware architecture search algorithm.
Experimental results show that, our DxPTA can find the appropriate PTA architectures for different transformer-based models (i.e., DeiT-T/S/B and BERT-B/L). 
It achieves up to 26mm$^2$ area, 4.8W power, 39mJ energy, and 6ms latency, for constraints of 50mm$^2$ area, 5W power, 50mJ energy, and 10ms latency; with 15.2x faster searching time than the exhaustive approach.
These results demonstrate the potential of DxPTA methodology for enabling efficient PTA designs for diverse AGI-based applications. 
\end{abstract}

\begin{IEEEkeywords}
Silicon Photonics, Photonic Transformer Accelerator (PTA), Design Space Exploration (DSE), Coherent Optical Dataflow, Hardware/Software (HW/SW) Co-Design.
\end{IEEEkeywords}

\section{Introduction}
\label{Sec_Intro}

\begin{figure}[t]
\centering
\includegraphics[width=\linewidth]{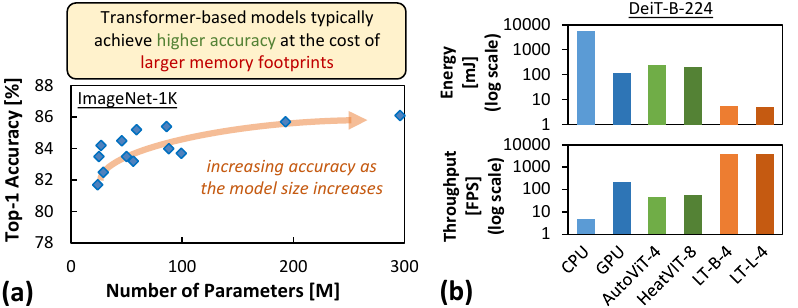}
\vspace{-0.6cm}
\caption{\textbf{(a)} Transformer-based models typically can improve the performance at the cost of larger memory size (i.e., higher number of parameters); based on the data from~\cite{Ref_Han_SurveyViT_TPAMI22}. 
\textbf{(b)} Experimental results of running Data-efficient Image Transformer Base (DeiT-B)~\cite{Ref_Touvron_Transformers_ICML21} with different compute platforms: CPU, GPU, CMOS-based accelerators (i.e., AutoViT-4bit~\cite{Ref_Li_AutoVitAcc_FPL22} and HeatViT-8bit~\cite{Ref_Dong_HeatViT_HPCA23}) and photonic-based Lightening-Transformer (LT) accelerators (i.e., LT-Base-4bit and LT-Large-4bit); based on data from~\cite{Ref_Zhu_LightTrans_HPCA24}.}
\label{Fig_Trends}
\vspace{-0.3cm}
\end{figure}

Transformer-based network models~\cite{Ref_Vaswani_Attention_NIPS17}, such as Vision Transformers (ViTs) and Large Language Models (LLMs), have emerged as prominent AI models with state-of-the-art performance for solving diverse machine learning tasks, such as vision and natural language processing (NLP)~\cite{Ref_Dosovitskiy_Transformers_ICLR21, Ref_Touvron_Transformers_ICML21, Ref_Khan_SurveyViT_CSUR22, Ref_Han_SurveyViT_TPAMI22}, hence potentially paving the way toward artificial general intelligence (AGI)~\cite{Ref_Mumuni_LLM4AGI_arXiv25}\cite{Ref_Yenduri_AGIsurvey_Access25}.
However, this state-of-the-art performance comes at high computational and memory requirements as shown in Fig.~\ref{Fig_Trends}(a), hence leading to huge power/energy consumption~\cite{Ref_Han_SurveyViT_TPAMI22}. 
This condition makes it difficult to obtain high performance efficiency when processing transformer models for wide-scale implementations across diverse applications.
A potential solution is employing specialized accelerators for expediting the inference of transformers, thus minimizing power consumption and improving energy efficiency~\cite{Ref_Lu_HW4MHA_SOCC20, Ref_Qi_AccelTrans_ICCAD21, Ref_Wang_Spatten_HPCA21, Ref_Sun_Vaqf_arXiv22, Ref_Zhou_Transpim_HPCA22, Ref_Jouppi_TPUv4_ISCA23, Ref_You_Vitcod_HPCA23}; see Fig.~\ref{Fig_Trends}(b).
However, conventional electronic accelerators face challenges related to their diminishing performance efficiency (i.e., increased power dissipation-per-unit area and slower performance gains), as transistor circuits reach the limits of Dennard scaling~\cite{Ref_Sunny_SurveyPhot4DL_JETC21}.

Recently, electronic-photonic integrated circuit (EPIC)-based solutions, so-called \textit{photonic accelerators}~\cite{Ref_Shiflett_Albireo_ISCA21, Ref_Shastri_Photonics4AI_NaturePhot21, Ref_Gu_LightInAI_TCASII22, Ref_Yin_Simphony_DAC25}, have been studied as an alternative for achieving significant speedup and efficiency improvements over the electronic accelerators, due to their ultra-high speed, high bandwidth, and low energy consumption~\cite{Ref_Gu_LightInAI_TCASII22}.
Therefore, employments of photonic accelerators for expediting neural network (NN) workloads are actively being explored. 
For instance, photonic tensor core (PTC) developments leveraging optical components such as Mach-Zehnder Interferometer (MZI)~\cite{Ref_Shen_DLwCoherentPhot_NaturePhot17}, Micro-Ring Resonator (MRR)-based bank~\cite{Ref_Tait_PhotWeightBanks_SciRep17}\cite{Ref_Sunny_CrossLight_DAC21}, and Phase Change Material (PCM)-based crossbar~\cite{Ref_Feldmann_PCMcrossbar_Nature21}. 
However, these works still target in accelerating traditional convolutional neural networks (CNN) workloads~\cite{Ref_Zhu_LightTrans_HPCA24}, hence indicating the need for further studies to enable high-performance and energy-efficient inference of transformer models. 
Therefore, the \textbf{targeted research problem} in this paper is \textit{how can we effectively enable high performance and energy-efficient inference of transformer models using photonic-based accelerators}? 
A solution to this problem may enable the efficient deployments of transformer-based models on photonic-based computing systems.

\subsection{State-of-the-art of Photonic Transformer Accelerators (PTAs) and Their Limitations}
\label{Sec_Intro_SOTA}

Recent works propose PTA designs that employ statically-operated PTCs using MRR banks~\cite{Ref_Li_SPRINT_TPDS22, Ref_Li_SPACX_HPCA22, Ref_Afifi_TRON_GLSVLSI23, Ref_Afifi_LLMoptical_GLSVLSI25, Ref_Afifi_ASTRA_TECS25} and PCM crossbars~\cite{Ref_Li_MERIT_TSUSC25}.
Another work proposes the \textit{Lightening-Transformer (LT)}~\cite{Ref_Zhu_LightTrans_HPCA24} based on dynamically-operated PTCs. 
It inspires further studies in digital-to-analog converter (DAC) design~\cite{Ref_Li_HyAtten_DATE25, Ref_Chang_PDAC_DAC25} and reconfigurability~\cite{Ref_Zhu_ENlighten_arXiv25}. 
LT improves the performance and efficiency of transformer inference over other PTAs by enabling dynamic operations of full-range input operands, making it the state-of-the-art PTA design. 
Despite their benefits, all these works still have the following limitations.
\begin{itemize}[leftmargin=*]
    \item Their architectures are developed without considering application constraints (e.g., area, power, energy, and latency), and hence their designs are not directly applicable for targeted applications and leading to sub-optimal performance and efficiency gains. 
    \item Their manual design approach needs a huge design time and power/energy consumption to develop a suitable architecture for the targeted applications, thus making this approach not scalable. 
\end{itemize}
To illustrate the limitations of state-of-the-arts and related research challenges, we perform an experimental case study, which will be discussed in Section~\ref{Sec_Intro_StudyChallenges}.

\subsection{Case Study and Research Challenges}
\label{Sec_Intro_StudyChallenges}

\begin{figure}[t]
\centering
\includegraphics[width=\linewidth]{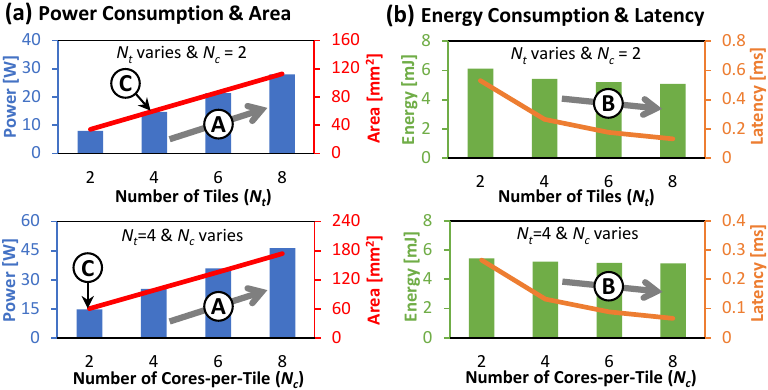}
\vspace{-0.7cm}
\caption{Experimental results considering different configurations of architecture parameters (i.e., $N_t$ and $N_c$) in the 4-bit LT accelerator for \textbf{(a)} power and area; and \textbf{(b)} energy consumption and latency considering the DeiT-Base~\cite{Ref_Touvron_Transformers_ICML21}.}
\label{Fig_CaseStudy}
\vspace{-0.5cm}
\end{figure}

We explore the impact of different architecture parameters of the state-of-the-art 4-bit LT accelerator~\cite{Ref_Zhu_LightTrans_HPCA24}. 
Here, we vary the number of tiles ($N_t$) and the number of cores-per-tile ($N_c$). 
For workload, we consider the DeiT-Base model~\cite{Ref_Touvron_Transformers_ICML21}.
Details of the LT hardware architecture and the experimental setup are provided in Section~\ref{Sec_Prelim_PTA} and Section~\ref{Sec_EvalMethod}, respectively.
Experimental results are shown in Fig.~\ref{Fig_CaseStudy}, from which we make the following key observations.
\begin{itemize}[leftmargin=*]
    \item Different configurations lead to different profiles of area, power, energy, and latency, thus highlighting the wide range of design choices for developing photonic accelerators.    
    \item Increasing $N_t$ or $N_c$ leads to higher power and larger area due to more complex circuitry (see \circled{A}), but it may reduce latency and energy consumption due to increased parallelism (see \circled{B}). 
    This shows the need for trade-off analysis in accelerator design. 
    \item The state-of-the-art LT design (with $N_t$=4 and $N_c$=2) may not meet the constraints. For instance, in low-power applications with max. 5W, the design incurs significantly more power ($\sim$15W) as shown by \circled{C}, indicating the need for a custom architecture.
\end{itemize}

These observations expose several research challenges in devising solutions for the targeted research problem, as outlined below.
\begin{itemize}[leftmargin=*]
    \item The solution should leverage the characteristics of photonic devices and optical dataflow to find the PTA architecture that meets all constraints, ensuring its applicability for diverse applications.
    \item The solution should minimize the searching time of PTA architecture, hence expediting the design time and providing a scalable design approach for diverse applications.  
\end{itemize}
    
\subsection{Our Novel Contributions}
\label{Sec_Intro_Novelty}

\begin{figure}[h]
\vspace{-0.3cm}
\centering
\includegraphics[width=\linewidth]{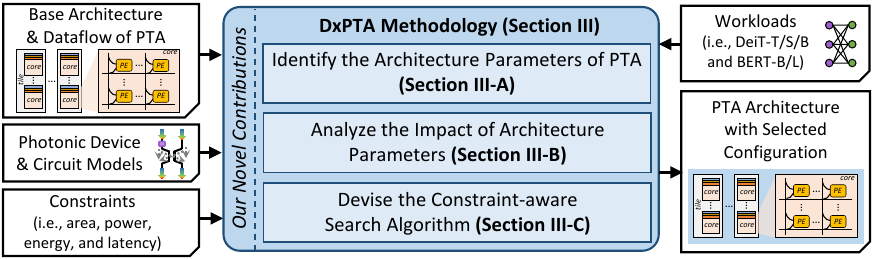}
\vspace{-0.7cm}
\caption{Our novel contributions in this work.}
\label{Fig_Novelty}
\vspace{-0.5cm}
\end{figure}

\begin{figure*}[t]
\centering
\includegraphics[width=\linewidth]{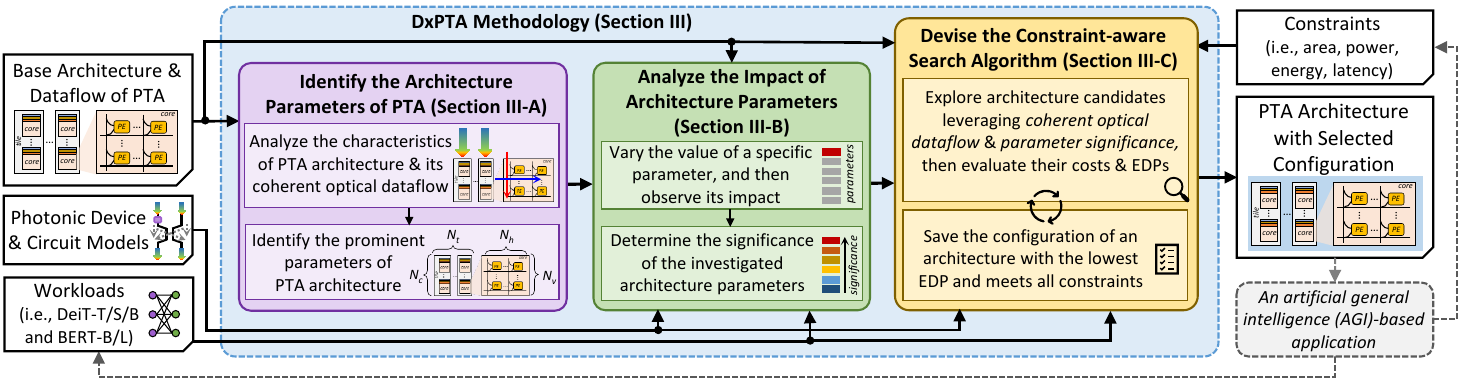}
\vspace{-0.7cm}
\caption{Our novel DxPTA methodology with its key steps: (1) identifying of the architecture parameters of PTA; (2) analyzing the impact of architecture parameters; and (3) devising the constraint-aware search algorithm.}
\label{Fig_DxPTA}
\vspace{-0.4cm}
\end{figure*}

To address the targeted research problem and related challenges, we propose \textit{\textbf{DxPTA}}, \textit{a novel architecture \underline{D}esign space e\underline{x}ploration methodology leveraging coherent optical dataflow-guided strategy for efficient hardware/software (HW/SW) co-design of \underline{P}hotonic \underline{T}ransformer \underline{A}ccelerators while meeting multiple constraints (i.e., area, power, energy, and latency)}.
It employs the following key steps (see an overview in Fig.~\ref{Fig_Novelty} and details in Fig.~\ref{Fig_DxPTA}).
\begin{itemize}[leftmargin=*]
    \item \textbf{Identify the architecture parameters of PTA (Section~\ref{Sec_DxDPTA_IdentifyArch}):} 
    It targets to analyze the characteristics of PTA architecture (including its hierarchy and photonic devices) and its coherent optical dataflow for identifying the prominent architecture parameters.
    \item \textbf{Analyze the impact of architecture parameters (Section~\ref{Sec_DxDPTA_ImpactArchParams}):} 
    It identifies the significance of architecture parameters (e.g., $N_t$ and $N_c$) by observing their impact on area, power, energy, and latency. The information will be leveraged for architecture search.
    \item \textbf{Devise the constraint-aware search algorithm (Section~\ref{Sec_DxDPTA_DeviseSearchAlg}):} 
    It explores architecture candidates by leveraging the coherent optical dataflow and parameter significance, evaluates their energy-delay products (EDPs), and then select the one that has the lowest EDP and meets all constraints.  
\end{itemize}

\textbf{Key Results:}
We evaluate our DxPTA methodology using Python implementation, and then run it on an Nvidia RTX 6000 Ada GPU machine, while considering diverse transformer workloads (i.e., DeiT-T/S/B\footnote{DeiT-T/S/B refers to DeiT-Tiny, DeiT-Small, and DeiT-Base, respectively.} and BERT-B/L\footnote{BERT-B/L refers to BERT-Base and BERT-Large, respectively.}). 
Experimental results show that, DxPTA successfully finds the accelerator architectures that meet all constraints. 
It achieves up to 26mm$^2$ area, 4.8W power, 39mJ energy, and 6ms latency across all investigated models, for constraints of 50mm$^2$ area, 5W power, 50mJ energy, and 10ms latency, with 15.2x faster searching time than the exhaustive approach.  

\section{Preliminaries}
\label{Sec_Prelim}

\subsection{Transformer-based Networks}
\label{Sec_Prelim_Transformers}

A transformer-based network typically consists of multiple identical blocks, known as \textit{encoder} and \textit{decoder} blocks.
Each block consists of a multi-head self-attention (MHA) module, a feed-forward network (FFN), shortcut connections, as well as a layer normalization (LN)~\cite{Ref_Zhu_LightTrans_HPCA24}. Furthermore, the decoder block also has cross-attention and masked self-attention modules. 
The basic encoder block can be formulated as Eq.~\ref{Eq_Encoder}-\ref{Eq_Encoder2}. Here, $\textbf{X}_l$ is the input sequences of $l$-th layer.
\begin{equation}
  \begin{split}
    \hat{\textbf{X}}_{l+1} = \textit{MHA}(\textit{LN}(\textbf{X}_l))+\textbf{X}_l
  \end{split}
  \label{Eq_Encoder}
\end{equation}
\begin{equation}
  \begin{split}
     \textbf{X}_{l+1} = \textit{FFN}(\textit{LN}(\hat{\textbf{X}}_{l+1}))+\hat{\textbf{X}}_{l+1} 
  \end{split}
  \label{Eq_Encoder2}
\end{equation}
Multi-head self-attention (MHA) module has $H$ self-attention heads, where each head transforms the input vector into separate vectors: query (\textbf{Q}), key (\textbf{K}), and value (\textbf{V}) vectors.
The attention function between these input vectors can be calculated using Eq.~\ref{Eq_Attention}. Here, $d_k$ is the dimension of \textbf{Q} and \textbf{K}. 
\begin{equation}
    Attention(\textbf{Q}, \textbf{K}, \textbf{V}) = softmax \left(\frac{\textbf{Q}\textbf{K}^{\intercal}}{\sqrt{d_k}}\right) \textbf{V}
  \label{Eq_Attention}
\end{equation}

\subsection{Photonic Transformer Accelerator (PTA)}
\label{Sec_Prelim_PTA}

\begin{figure}[t]
\centering
\includegraphics[width=\linewidth]{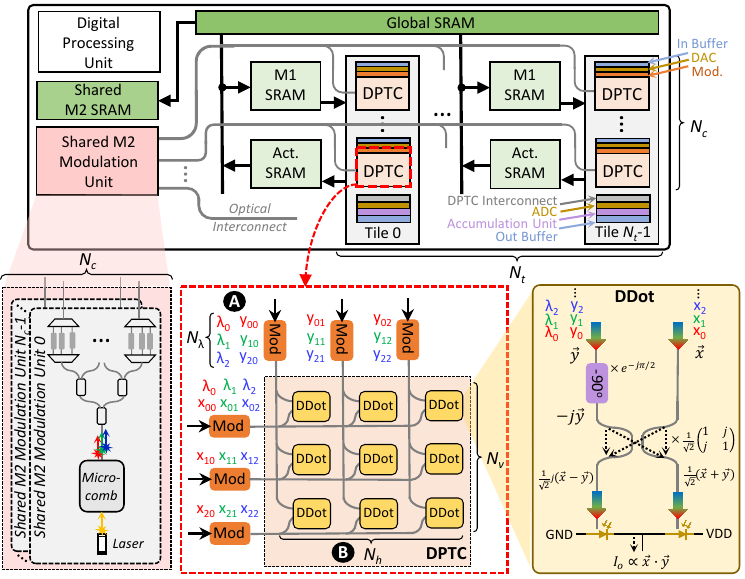}
\vspace{-0.7cm}
\caption{Architecture of the LT accelerator; based on ~\cite{Ref_Zhu_LightTrans_HPCA24}.}
\label{Fig_LTarch}
\vspace{-0.5cm}
\end{figure}

In this work, \textit{we focus on the LT accelerator~\cite{Ref_Zhu_LightTrans_HPCA24} as the reference design since it is the state-of-the-art PTA architecture}, whose descriptions are provided in the following; see an overview in Fig.~\ref{Fig_LTarch}. 
The LT Accelerator consists of analog photonic computing elements for accelerating general matrix multiplication (GEMM), optical interconnect for data transmission, and electronics for other operations (e.g., data storage, signal conversion, nonlinear functions, and softmax); see Fig.~\ref{Fig_LTarch}. 
Following are its key design points.
\begin{itemize}[leftmargin=*]
    \item A single LT chip typically contains $N_t$ tiles, and each tile clusters $N_c$ \textit{dynamically-operated photonic tensor cores (DPTCs)}. 
    \item A DPTC is known as the \textit{core} of LT accelerator, and it contains an array of $N_h$$\times$$N_v$ \textit{dynamically-operated dot-product engine (DDot)}. 
    \item A single DDot performs optical dot-product operation between two full-range vectors $\vec{x}$ and $\vec{y}$ based on coherent interference. 
\end{itemize}
Dot-product operation in DDot is performed with the following steps.
First, each pair of inputs ($x_i,y_i$) is encoded in the same wavelength $\lambda_i$ through the \textit{wavelength-division multiplexing (WDM)} technique. 
These signals are passed to the two arms of 50:50 \textit{directional coupler (DC)} with -90$\degree$ \textit{phase shifter (PS)}.
The outputs of DC ($z^0_i,z^1_i$) are orthogonal in complex plane and can be computed with Eq.~\ref{Eq_DC}. 
\begin{equation}
  \begin{split}
    \begin{pmatrix}
        z_i^{0} \\
        z_i^{1}
    \end{pmatrix}
    & =
    \underbrace{
    \frac{1}{\sqrt{2}}
    \begin{pmatrix}
        1 & j \\
        j & 1
    \end{pmatrix}
    }_{DC}
    \underbrace{
    \begin{pmatrix}
        1 & 0 \\
        0 & e^{-j\pi/2}
    \end{pmatrix}
    }_{PS}
    \begin{pmatrix}
        x_i \\
        y_i
    \end{pmatrix} \\
    & = \frac{1}{\sqrt{2}}
    \begin{pmatrix}
        x_i + y_i \\
        j(x_i - y_i)
    \end{pmatrix}
  \end{split}
  \label{Eq_DC}
\end{equation}
The photodiode (PD) at each output port of DC, converts the signals into photocurrent, which is proportional to the accumulated optical intensities of the input signals. 
Hence, the output current ($I_0$) follows the relation of $I_0 \propto \vec{x} \cdot \vec{y}$.   

\begin{figure}[t]
\centering
\includegraphics[width=\linewidth]{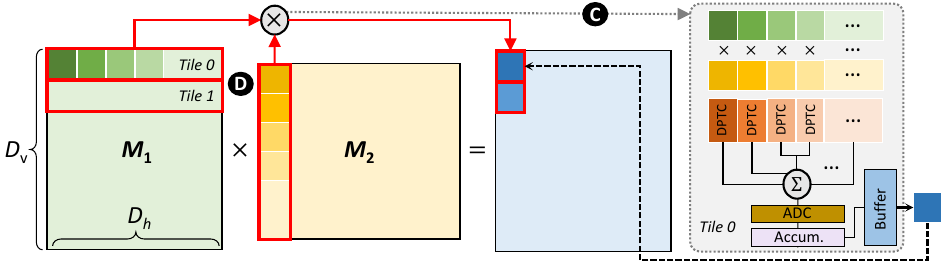}
\vspace{-0.7cm}
\caption{Data tiling mechanism in the accelerator~\cite{Ref_Zhu_LightTrans_HPCA24}, which partitions data from matrix $M_1$ along the $D_v$ dimension and map them to different tiles.}
\label{Fig_Tiling}
\vspace{-0.4cm}
\end{figure}

\vspace{-0.1cm}
\section{Our DxPTA Methodology}
\label{Sec_DxDPTA}

This methodology identifies the architecture parameters, analyzes the impact of parameters, and develops the constraint-aware search algorithm; which are further described below (an overview in Fig.~\ref{Fig_DxPTA}).

\begin{figure*}[t]
\centering
\includegraphics[width=\linewidth]{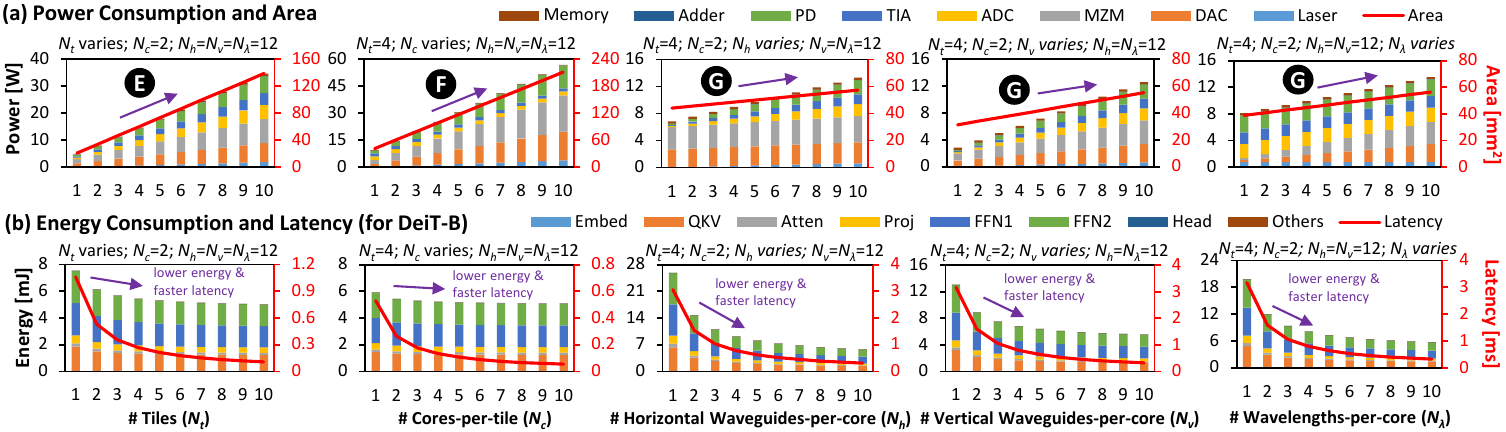}
\vspace{-0.6cm}
\caption{\textbf{(a)} Power and area for different configurations of parameters. \textbf{(b)} Energy consumption and latency for different configurations of parameters when running DeiT-B.
We observe similar trends for different workloads (DeiT-T/S and BERT-B/L). }
\label{Fig_ImpactParams}
\vspace{-0.2cm}
\end{figure*}

\subsection{Identifying the Architecture Parameters}
\label{Sec_DxDPTA_IdentifyArch}
\vspace{-0.2cm}

Discussion in Section~\ref{Sec_Intro_StudyChallenges} suggests that the configuration of architecture parameters is important for determining the performance and efficiency of the PTA. 
Therefore, \textit{this step aims to identify parameters that should be customized when designing the PTA}.
To achieve this, we first analyze the hierarchy of the PTA base architecture and its coherent optical dataflow, and make the following key observations.
\begin{itemize}[leftmargin=*]  
    \item Combining the dynamically-operated DDots and the coherent optical dataflow enables multi-wavelength processing which maximizes spectral parallelism and throughput. 
    From such coherent dataflow and operations, we observe some characteristics below. 
    \begin{itemize}
        \item Multiple wavelengths can be processed in the same DDot unit without requiring prior data programming; see~\circledB{A} in Fig.~\ref{Fig_LTarch}. 
        \item An $N_h$$\times$$N_v$ DDot array is employed to perform multiplications, which defines the parallelism level in a core; see~\circledB{B} in Fig.~\ref{Fig_LTarch}.  
        \item Multiple cores can be employed to process a chunk of data, which defines the parallelism level in a tile; see~\circledB{C} in Fig.~\ref{Fig_Tiling}. 
        \item Multiple tiles can be employed to process multiple data chunks, which defines the parallelism level in a chip; see~\circledB{D} in Fig.~\ref{Fig_Tiling}.  
    \end{itemize}
    \item On-chip global SRAM should have the minimum required size for holding the largest activations in a layer due to layer-by-layer processing, and buffering a portion of off-chip data based on the tiling approach (see Fig.~\ref{Fig_Tiling}). 
    Hence, global SRAM size should not be reduced below or increased significantly over this minimum required size, as this will increase the expensive off-chip data access (i.e., high access latency and access energy) or aggravate the static power, respectively~\cite{Ref_Putra_DRMap_DAC20, Ref_Putra_ROMANet_TVLSI21, Ref_Putra_PENDRAM_arXiv24}.
\end{itemize}
These observations expose the parameters that should be configured for developing an appropriate accelerator architecture, i.e., \textit{number of tiles ($N_t$)}, \textit{number of cores-per-tile ($N_c$)}, \textit{number of input horizontal waveguides-per-core ($N_h$)}, \textit{number of input vertical waveguides-per-core ($N_v$)}, and \textit{number of wavelengths ($N_{\lambda}$)}.

\subsection{Analyzing the Impact of Parameters}
\label{Sec_DxDPTA_ImpactArchParams}

To ensure the accelerator meets the constraints, an appropriate configuration of parameters (i.e., $N_t$, $N_c$, $N_h$, $N_v$, and $N_{\lambda}$) is needed.
A promising solution is employing design space exploration (DSE). 
However, the design space is large due to a high number of possible configurations from different parameter sizes, indicating the need for an optimization.
Hence, \textit{this step aims to identify the significance of parameters, which will be used to efficiently guide the DSE process}. 
To achieve this, we propose to employ the following steps.
\begin{itemize}[leftmargin=*]
    \item We perform an experimental case study that varies the value of a specific parameter and analyze its impact on different metrics, i.e., area, power, energy, and latency.
    \item Afterward, we evaluate the significance score ($S$) for each parameter on a specific metric using Eq.~\ref{Eq_Score}, whose mechanism is also presented as pseudocode in Alg.~\ref{Alg_ObserveImpact}. 
    Here, $s_i$ is the ratio between the values of metric-$m$ (i.e., area $A$ or power $P$) from architecture with $i$+1 units and architecture with $i$ units, which represents the impact of unit addition; while $K$ is the total number of ratios. 
    \begin{equation}
      \begin{split}
        S = &  \frac{1}{K} \sum_{i=1}^{K} s_i = \frac{1}{K} \sum_{i=1}^{K} \frac{m_{i+1\text{ units}}}{m_{i\text{ units}}} \\ 
        & \text{with}  \;\;\; m \in \{A, P\}
      \end{split}
      \label{Eq_Score}
    \end{equation}
\end{itemize}

Experimental results are shown in Fig.~\ref{Fig_ImpactParams}, from which we make the following key observations.
\begin{itemize}[leftmargin=*]
    \item Increasing parameter size leads to higher power and area, but potentially reduces latency and energy due to higher parallelism.  
    \item $N_t$ has the highest impact as each additional tile leads to the highest significance $S$ with 1.26x higher power and 1.24x larger area on average; see \circledB{E}. 
    Introducing a new tile means addition of single/multiple cores and peripherals (e.g., DAC and ADC). 
    \item $N_c$ has relatively high impact as each additional core leads to high significance $S$ with 1.23x higher power and 1.20x larger area on average; see \circledB{F}. 
    Introducing a new core means addition of a DDot array with increased size of peripherals (e.g., accumulator). 
    \item $N_v$, $N_h$, or $N_{\lambda}$ have comparable impact to each other, but they are lower than $N_t$ and $N_c$. 
    For each additional unit of $N_v$, $N_h$, or $N_{\lambda}$, power and area are increased by up to 1.16x and 1.06x, respectively; see \circledB{G}. 
    Introducing new DDots within the core or new wavelengths only slightly increases circuit complexity and size (e.g., broadcast unit in the array).  
\end{itemize}

\begin{algorithm}[t]
\small
\caption{Observing the significance of architecture parameters}
\label{Alg_ObserveImpact}
\begin{algorithmic}[1]
\renewcommand{\algorithmicrequire}{\textbf{INPUT:}}
\renewcommand{\algorithmicensure}{\textbf{OUTPUT:}}
\REQUIRE Maximum number of observations ($J$=10); \\
\ENSURE Significance score of each parameter on area ($S_A$) and power ($S_P$); \\
\smallskip
\textbf{BEGIN} \\
\textbf{Process:} \\  
\FOR{each investigated parameter}
  \STATE $N_t$ = 4; $N_c$ = 2; $N_v$ = 12; $N_h$ = 12; $N_{\lambda}$ = 12; \textcolor{teal}{// init default values} \\
  \FOR{($j$=1; $j$$<$($J$+1); $j$++)}
    \STATE set the investigated parameter value with $j$;
    \STATE $cfg[j]$ = construct($N_t$, $N_c$, $N_v$, $N_h$,$N_{\lambda}$); \\
    \STATE $A[j]$, $P[j]$ = eval\_hw($cfg[j]$);
    \STATE $i$ = $j$-1;
    \IF{($i$$>$0)}
      \STATE $s_A[i]$ = $A[i$+$1]/A[i]$; \textcolor{teal}{// compute ratio for area} \\
      \STATE $s_P[i]$ = $P[i$+$1]/P[i]$; \textcolor{teal}{// compute ratio for power} \\
    \ENDIF
  \ENDFOR
  \STATE $K$ = $J$-1;
  \STATE $S_A$ = $\frac{1}{K} \sum_{i=1}^{K} s_A[i]$; \textcolor{teal}{// compute significance score for area with Eq.~\ref{Eq_Score}} \\
  \STATE $S_P$ = $\frac{1}{K} \sum_{i=1}^{K} s_P[i]$; \textcolor{teal}{// compute significance score for power with Eq.~\ref{Eq_Score}} \\
\ENDFOR
\STATE \textbf{return} $S_A$, $S_P$; \\
\textbf{END}
\end{algorithmic}
\end{algorithm}
\setlength{\textfloatsep}{6pt}

\vspace{-0.4cm}
\subsection{The Constraint-aware Search Algorithm}
\label{Sec_DxDPTA_DeviseSearchAlg}

Based on observations in Section~\ref{Sec_DxDPTA_ImpactArchParams}, we develop the following optimization strategy for DSE process.
\begin{itemize}[leftmargin=*]
    \item Exploring $N_t$ and $N_c$ should be performed carefully since their slight changes may incur significant changes on area and power. 
    \item Exploring $N_v$, $N_h$, and $N_c$ may be performed more aggressively than $N_t$ and $N_c$, since their slight changes do not incur significant changes on area and power consumption.  
    \item The data dimension is typically evenly sized across layers, which should be leveraged to maximize the resource utilization by selecting the evenly-sized dimension for architecture parameters. 
\end{itemize}
We leverage this strategy for devising \textit{a constraint-aware search algorithm}.
Its key ideas are shown in Alg.~\ref{Alg_Search} and discussed below. 
\begin{itemize}[leftmargin=*]
    \item We determine a set of values for each parameter, which defines its search space; see Alg.~\ref{Alg_Search}: lines 3-10.
    It is stored in $T_{cnd}$, $C_{cnd}$, $V_{cnd}$, $H_{cnd}$, $G_{cnd}$ for $N_t$, $N_c$, $N_v$, $N_h$, $N_{\lambda}$, respectively. 
    The search spaces for $T_{cnd}$ and $C_{cnd}$ employ incremental values, while the ones for $V_{cnd}$, $H_{cnd}$ and $G_{cnd}$ are optimized using progressive values with exploration $step$ based on evenly-sized data dimension. 
    \item Then, each configuration candidate for the architecture is explored.
    It is performed by investigating each combination of values from different parameters, and evaluating the area, power, energy, and latency profiles; see Alg.~\ref{Alg_Search}: lines 11-14.
    \item We evaluate if these profiles meet all constraints and if the energy-delay product (EDP) is lower than the recorded one.
    If so, then the configuration is saved; see Alg.~\ref{Alg_Search}: lines 15-22.
    Here, EDP is the metric for reflecting both performance and energy efficiency.
    \item When the DSE process is finished, the last recorded configuration ($cfg_{svd}$) is selected as the final solution; see Alg.~\ref{Alg_Search}: line 23.
\end{itemize}

\begin{algorithm}[t]
\small
\caption{Our constraint-aware search algorithm}
\label{Alg_Search}
\begin{algorithmic}[1]
\renewcommand{\algorithmicrequire}{\textbf{INPUT:}}
\renewcommand{\algorithmicensure}{\textbf{OUTPUT:}}
\REQUIRE \textbf{(1)} Maximum number of parameter sizes ($N_z$=12); \\
\textbf{(2)} Progressive exploration step for non-significant parameters ($step$=2); \\
\textbf{(3)} Targeted pre-trained network model ($net$);
\textbf{(4)} Constraints for area ($const_A$), power ($const_P$), energy ($const_E$), and latency ($const_L$);
\ENSURE \textbf{(1)} Final configuration of the architecture parameters ($cfg_{svd}$); \\
\smallskip
\textbf{BEGIN} \\
\smallskip
\textbf{Initialization:} \\
\STATE $Z_{1}$ = []; $Z_{2}$ = []; \\
\STATE $EDP_{svd}$ = 1000; \\
\smallskip
\textbf{Process:} \\  
\smallskip
\underline{// Define the search space for each parameter}
\smallskip
\FOR{($n_z$=1; $n_z$$<$($N_z$+1); $n_z$++)}
  \STATE $Z_{1}$ = append($Z_{1}$, $n_z$);
  \IF{($n_z$ mod $step$ == 0)}
    \STATE $Z_{2}$ = append($Z_{2}$, $n_z$);
  \ENDIF
\ENDFOR
\STATE $T_{cnd}$ = $Z_{1}$; $C_{cnd}$ = $Z_{1}$; \\
\STATE $V_{cnd}$ = $Z_{2}$; $H_{cnd}$ = $Z_{2}$; $G_{cnd}$ = $Z_{2}$; \\
\smallskip
\underline{// Construct and evaluate the configurations} 
\smallskip
\FOR{each combination from ($\forall$ $n_t$ $\in$ $T_{cnd}$), 
       ($\forall$ $n_c$ $\in$ $C_{cnd}$),
       ($\forall$ $n_v$ $\in$ $V_{cnd}$),
       ($\forall$ $n_h$ $\in$ $H_{cnd}$), and
       ($\forall$ $n_{\lambda}$ $\in$ $G_{cnd}$)}
  \STATE $cfg_{cnd}$ = construct($N_t$, $N_c$, $N_v$, $N_h$, $N_{\lambda}$);
  \STATE $A_{cnd}$, $P_{cnd}$ = eval\_hw($cfg_{cnd}$); \textcolor{teal}{// evaluate area \textit{A} and power \textit{P}} \\
  \STATE $E_{cnd}$, $L_{cnd}$ = eval\_wload($cfg_{cnd}$, $net$); \textcolor{teal}{// evaluate energy \textit{E} and latency \textit{L}} \\
  \IF{($A_{cnd}$ $<$ $const_A$) and ($P_{cnd}$ $<$ $const_P$) and \\
      ($E_{cnd}$ $<$ $const_E$) and ($L_{cnd}$ $<$ $const_L$)}
      \STATE $EDP_{cnd}$ = calc\_EDP ($E_{cnd}$, $L_{cnd}$); \textcolor{teal}{// evaluate EDP} \\
      \IF{($EDP_{cnd}$ $<$ $EDP_{svd}$)} 
        \STATE $EDP_{svd}$ = $EDP_{cnd}$;
        \STATE $cfg_{svd}$ = $cfg_{cnd}$;
    \ENDIF
  \ENDIF
\ENDFOR
\STATE \textbf{return} $cfg_{svd}$; \textcolor{teal}{// final $N_t$, $N_c$, $N_v$, $N_h$, and $N_{\lambda}$} \\
\textbf{END}
\end{algorithmic}
\end{algorithm}
\setlength{\textfloatsep}{6pt}

\vspace{-0.4cm}
\section{Evaluation Methodology}
\label{Sec_EvalMethod}

\begin{figure}[t]
\centering
\includegraphics[width=\linewidth]{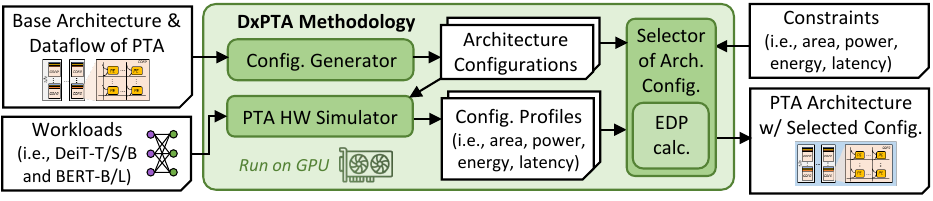}
\vspace{-0.7cm}
\caption{Experimental setup used in this work.}
\label{Fig_ExpSetup}
\end{figure}

We implement the DxPTA methodology using PyTorch and run it on the Nvidia RTX 6000 Ada GPU machine; see Fig.~\ref{Fig_ExpSetup}.
For hardware evaluation, we employ the state-of-the-art PTA hardware simulator~\cite{Ref_Zhu_LightTrans_HPCA24} with 4-bit precision that has been evaluated using the Lumerical Interconnect tools, and incorporate it into the DxPTA.
For workloads, we use DeiT-T, DeiT-S, DeiT-B, BERT-B, and BERT-L models.
For comparison partners, we consider the LT accelerators (i.e., LT-Base and LT-Large)~\cite{Ref_Zhu_LightTrans_HPCA24} as the state-of-the-art PTA designs, and an exhaustive approach as the search technique.
For constraints, we consider 50mm$^2$ area, 5W power, 50mJ energy, and 10ms latency, to show the applicability of DxPTA for providing solutions under any application requirements.
Evaluation metrics include area, power, energy, latency, and search time.

\section{Experimental Results and Discussion}
\label{Sec_Results}

\subsection{Ensuring the PTA Architecture Design to Meet All Design Constraints}
\label{Sec_Results_MeetConstraints}

\begin{figure*}[t]
\centering
\includegraphics[width=\linewidth]{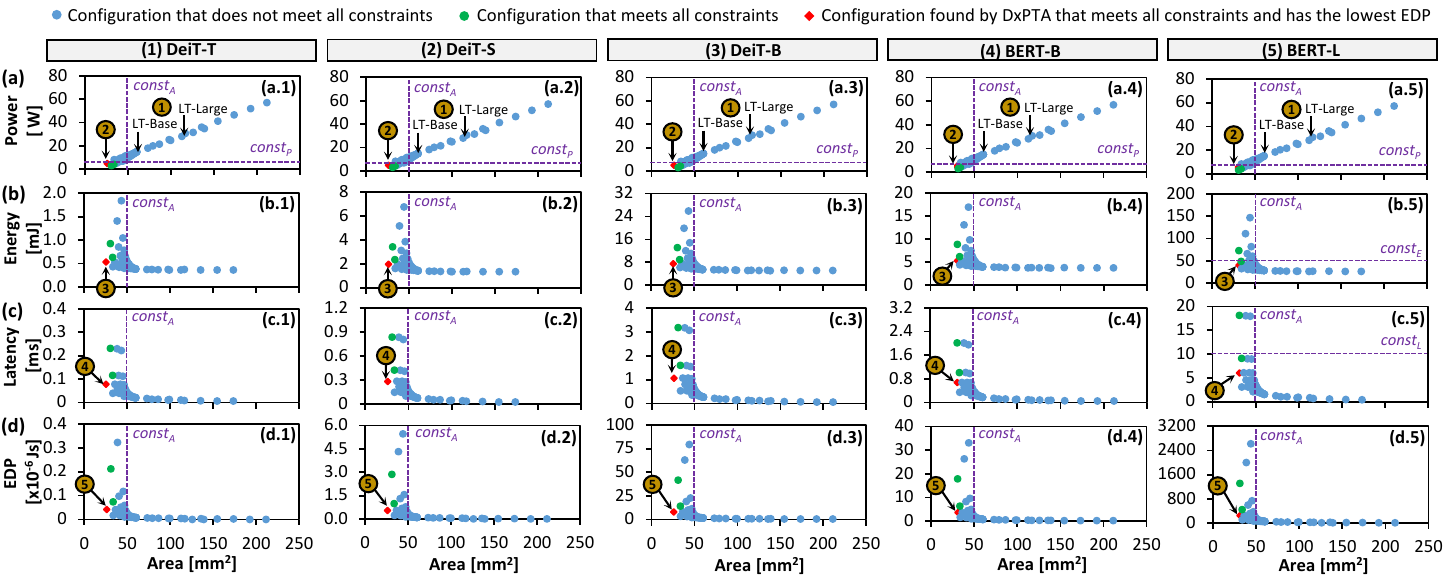}
\vspace{-0.7cm}
\caption{Experimental results of \textbf{(a)} power vs. area, \textbf{(b)} energy vs. area, \textbf{(c)} latency vs. area, and \textbf{(d)} EDP vs. area for different samples of configurations during DSE process, across different workloads: DeiT-T, DeiT-S, DeiT-B, BERT-B, and BERT-L.}
\label{Fig_Results}
\vspace{-0.2cm}
\end{figure*}

\begin{figure}[t]
\centering
\includegraphics[width=\linewidth]{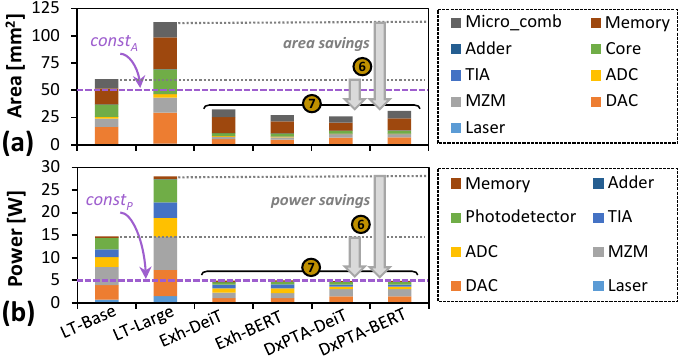}
\vspace{-0.7cm}
\caption{Experimental results of \textbf{(a)} area and \textbf{(b)} power for LT-Base, LT-Large, exhaustive search-based accelerators for DeiT/BERT models (i.e., Exh-DeiT/BERT), and DxPTA-based accelerators for DeiT/BERT models (i.e., DxPTA-DeiT/BERT).}
\label{Fig_Results_HW_AnP}
\end{figure}

Fig.~\ref{Fig_Results} provides experimental results of area, power, energy consumption, and latency for different samples of investigated architecture configurations during DSE process across different workloads. 
These results show that, the state-of-the-art accelerators (i.e., LT-Base and LT-Large) do not meet all constraints at once, as they incur significantly larger area and higher power than the respective constraints; see \tikzmarknode[circledColor,draw=black,fill=brown,text=black]{t1}{1}.
Specifically, LT-Base and LT-Large incur about 60mm$^2$ and 112mm$^2$, respectively ($>$50mm$^2$ constraint).
These sizes are dominated by memory, DAC, and cores; see Fig.~\ref{Fig_Results_HW_AnP}(a). 
In terms of power, LT-Base and LT-Large incur about 15W and 28W power, respectively ($>$5W constraint).
These power values are dominated by Mach-Zender Modulator (MZM), DAC, photodetector, and ADC; see Fig.~\ref{Fig_Results_HW_AnP}(b).
The reason is that, LT-Base and LT-Large designs have fixed configurations for accelerating diverse workloads, thus they may not be applicable for different requirements.

In contrast, our DxPTA consistently finds the suitable configurations that meet all constraints and have the lowest EDP scores among the candidates across different workloads; see \tikzmarknode[circledColor,draw=black,fill=brown,text=black]{t1}{2} for power and area, \tikzmarknode[circledColor,draw=black,fill=brown,text=black]{t1}{3} for energy consumption, \tikzmarknode[circledColor,draw=black,fill=brown,text=black]{t1}{4} for latency, and \tikzmarknode[circledColor,draw=black,fill=brown,text=black]{t1}{5} for EDP. 
The reason is that, DxPTA employs a search algorithm that incorporates all constraints in its exploration process, ensuring the selected configuration to fulfills the requirements.
Consequently, the DxPTA-generated accelerators significantly reduce the area and power as compared to LT-Base and LT-Large, by achieving up to 76.9\% area saving and 82.7\% power saving; see \tikzmarknode[circledColor,draw=black,fill=brown,text=black]{t1}{6} in Fig.~\ref{Fig_Results_HW_AnP}.
Furthermore, our DxPTA-generated accelerators also achieve comparable area and power consumption to the accelerators whose configurations generated from exhaustive search (i.e., Exh-DeiT and Exh-BERT); see \tikzmarknode[circledColor,draw=black,fill=brown,text=black]{t1}{7} in Fig.~\ref{Fig_Results_HW_AnP}. 
The reason is that, DxPTA already considers the significance of parameters in its search strategy to ensure the coverage of potential configuration candidates in the DSE process. 
Hence, DxPTA can find configurations that are close to the ones from the exhaustive search.

\subsection{Enabling High-Performance and Energy-Efficient Transformer Inference}
\label{Sec_Results_PerformEfficiency}

\begin{figure}[t]
\centering
\includegraphics[width=\linewidth]{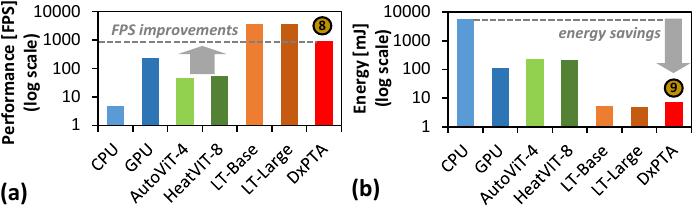}
\vspace{-0.7cm}
\caption{Experimental results of \textbf{(a)} performance and \textbf{(b)} energy consumption of DeiT-B processing using CPU, GPU, electronic accelerators (i.e., AutoViT-4b~\cite{Ref_Li_AutoVitAcc_FPL22} and HeatViT-8b~\cite{Ref_Dong_HeatViT_HPCA23}), state-of-the-art PTAs (i.e., LT-Base and LT-Large~\cite{Ref_Zhu_LightTrans_HPCA24}), and DxPTA.}
\label{Fig_Results_HW_PnE}
\end{figure}

Fig.~\ref{Fig_Results_HW_PnE} presents the performance (i.e., FPS: frame-per-second) and energy consumption of DeiT-B processing using different platforms: CPU, GPU, electronic accelerators, state-of-the-art PTAs, and our DxPTA-based PTA.
These results show that, the DxPTA-based PTA achieves comparable performance (FPS) to the LT-Base and LT-Large designs, and provides significant improvements from conventional platforms; see \tikzmarknode[circledColor,draw=black,fill=brown,text=black]{t1}{8}.
Specifically, the DxPTA-based PTA improves the performance by 189x from CPU, 4.1x from GPU, 20.1x from AutoVit-4, and 17.2x from HeatVIT-8. 
Such remarkable performance improvements come from the high-speed nature of light propagation.
The DxPTA-based PTA also achieves comparable energy-efficiency to the LT-Base and LT-Large designs, and provides significant energy savings from conventional platforms; see \tikzmarknode[circledColor,draw=black,fill=brown,text=black]{t1}{9}.
Specifically, the DxPTA-based PTA saves energy consumption by 782.1x from CPU, 15.2x from GPU, 31.6x from AutoVit-4, and 27.6x from HeatVIT-8.
Such energy efficiency improvements come from its lightweight optical-based processing and minimum energy consumption from electronic parts. 
Note, all these improvements are achieved by DxPTA while meeting all given constraints at once, further highlighting the benefits of our DxPTA methodology.

\subsection{Architecture Searching Time Speedup}
\label{Sec_Results_SearchTime}

\begin{figure}[t]
\centering
\includegraphics[width=\linewidth]{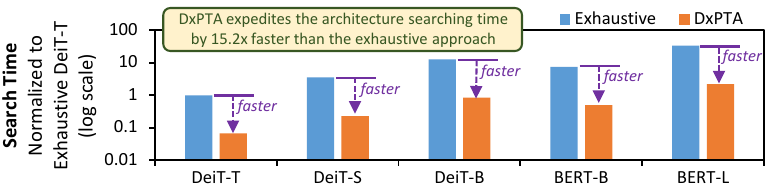}
\vspace{-0.7cm}
\caption{Searching time profiles of the exhaustive search approach and the searching strategy in our DxPTA.}
\label{Fig_Results_SearchTime}
\end{figure}

Fig.~\ref{Fig_Results_SearchTime} presents the searching time of the exhaustive approach and the guided search in our DxPTA across different workloads.
These results highlight that DxPTA achieves a significant searching time speedup, i.e., by 15.2x faster than the exhaustive one.
This speedup comes from the optimized search space in DxPTA through exploration steps, guided by data dimension, coherent optical dataflow, and parameter significance. 
This speedup is beneficial to make the DxPTA methodology an efficient and scalable solution for developing appropriate PTA architectures under different possible design constraints. 

\section{Conclusion}
\label{Sec_Conclusion}

We propose a novel DxPTA methodology to perform DSE for enabling efficient HW/SW co-design of the PTA architecture that meets all given constraints. 
DxPTA identifies the prominent architecture parameters based on the coherent optical dataflow, analyzes the significance of parameters, and then devises a constraint-aware search algorithm.
Experiments show that, DxPTA successfully finds the appropriate PTA architectures for different transformer models, achieving up to 26mm$^2$ area, 4.8W power, 39mJ energy, and 6ms latency, for constraints of 50mm$^2$ area, 5W power, 50mJ energy, and 10ms latency; with 15.2x faster search time than the exhaustive approach.
These results demonstrate the potential of DxPTA methodology for enabling efficient PTA design automation for diverse AGI-based applications.

% ...

\bibliographystyle{IEEEtran}
\bibliography{bibliography}

\end{document}